\documentclass[12pt,preprint]{aastex}
\usepackage{graphicx}
\usepackage{amssymb}
\footnotesize 
\newdimen\digitwidth 
\setbox0=\hbox{\rm0} 
\digitwidth=\wd0
\catcode`!=\active 
\def!{\kern\digitwidth} 
\normalsize

\begin{document}
\title{An improved solar wind electron-density model for pulsar
timing} 
\author{X. P. You\altaffilmark{1,2,3},
G.  B. Hobbs\altaffilmark{2}, W. A. Coles\altaffilmark{4}, 
R. N. Manchester \altaffilmark{2}, J. L. Han\altaffilmark{1}}
\altaffiltext{1}{National Astronomical Observatories, Chinese Academy
of Sciences, Beijing 100012, China. Email:xpyou@ns.bao.ac.cn} 
\altaffiltext{2}{Australia
Telescope National Facility, CSIRO, PO~Box~76, Epping, NSW~1710,
Australia} 
\altaffiltext{3}{Present address: School of Physics, Southwest University,
Chongqing 400175, China} 
\altaffiltext{4}{Electrical and Computer
Engineering, University of California at San Diego, La Jolla,
California 92093, USA}

\begin{abstract}
 Variations in the solar wind density introduce variable delays into
 pulsar timing observations.  Current pulsar timing analysis programs
 only implement simple models of the solar wind, which not only limit
 the timing accuracy, but can also affect measurements of pulsar
 rotational, astrometric and orbital parameters.  We describe a new
 model of the solar wind electron density content which uses
 observations from the Wilcox Solar Observatory of the solar magnetic
 field. We have implemented this model into the \textsc{tempo2} pulsar
 timing package.  We show that this model is more accurate than
 previous models and that these corrections are necessary for high
 precision pulsar timing applications.
\end{abstract}


\section{Introduction}
It is now possible to make timing observations of millisecond pulsars
to a precision of $\sim$100\,ns.  One of the most exciting
applications of such data-sets is to search for the signatures of
gravitational waves passing over the Earth.  This is a major goal of the
Parkes Pulsar Timing Array (PPTA) project \citep{hob05,man06}, which
aims to observe 20 millisecond pulsars with a timing precision close
to 100\,ns over more than five years. Many phenomena can affect the
pulse arrival times at this level of timing precision. A major
contributor at our primary observing frequency of $\sim$1400\,MHz is
the interstellar and interplanetary medium \citep{yhc+07}.  A small
change in a pulsar's dispersion measure (DM; the integrated electron
density along the line of sight to the pulsar) can cause significant
time delays in the pulse arrival times.  For example, at an observing
frequency of 1400\,MHz, a time delay of 100\,ns is caused by a DM
variation of only $\sim 5\times10^{-5}$\,cm$^{-3}$\,pc. At this level,
the solar wind effect is significant when the line of sight to the
pulsar passes within $\sim$60$^\circ$ of the Sun.

The standard pulsar timing programs
(\textsc{tempo1}\footnote{http://www.atnf.csiro.au/research/pulsar/tempo}
and \textsc{tempo2}; see Hobbs, Edwards \& Manchester
2006\nocite{hem06}, Edwards, Hobbs \& Manchester 2006\nocite{ehm06})
calculate the solar contribution, ${\rm DM}_\odot$, from a spherically
symmetric model of the solar wind density which assumes a quadratic
decrease with solar distance and ignores temporal variation:
\begin{equation}\label{eqn:model}
  {\rm DM}_\odot = 4.85\times10^{-6}\,n_0\,\frac{\theta}{\sin\theta}\;{\rm cm^{-3}pc},
\end{equation}
where $n_0$ is the electron density at $1$~AU from the Sun (in
cm$^{-3}$) and $\theta$ is the pulsar-Sun-observatory angle.  By
default, \textsc{tempo2} chooses $n_0 = 4$~cm$^{-3}$ whereas
\textsc{tempo1} uses $n_0 =10$~cm$^{-3}$. However, the true electron
density of the solar wind can change with longitude, latitude and time
by a factor of at least four \citep{mbf+00}.  \citet{yhc+07}
demonstrated that this simple model is inadequate for PSR~J1022+1001,
a pulsar that lies close to the ecliptic plane.

There have been several previous analyses of the timing delays or DM
variations due to the solar wind that occur in pulsar timing
observations. For instance, the ecliptic latitude ($\beta$) of the
Crab pulsar is only $-1.29^{\circ}$. \citet{lps88} showed, using a few
observations within $5^{\circ}$ of the Sun, that the maximum time
delay due to the solar wind was about 500\,$\mu$s at 610
MHz. Similarly, \citet{pw91} showed that the DM changed by $\sim
0.002\,{\rm pc\,cm}^{-3}$ when the line of sight to PSR~B0950+08
($\beta = -4.62^\circ$) is close to the Sun. \citet{cbl+96} observed
PSR~B1821$-$24 between the years 1989 and 1993 and showed that between
December and January each year their timing residuals were
significantly affected by the solar corona. More recently,
\citet{sns+05} and \citet{lkn+06} analysed data of PSRs~J1713+0747 and
J0030+0451 using the \textsc{tempo1} model, but, instead of holding
the electron density at 1\,AU fixed, they fitted for this scaling
factor.  They obtained that $n_0 = 5\pm4$\,cm$^{-3}$ and $n_0 =
7\pm2$\,cm$^{-3}$ respectively.

\citet{sfal97} argued that the planetary companions to PSR~B1257+12
\citep{wol94} were artefacts of incorrectly modelling the solar wind.
The closest planet to the pulsar produces a 25.3\,d periodicity in the
timing residuals which is remarkably close to periodicities seen in
Pioneer 10 spacecraft data which are thought to be due to patterns in
the solar wind caused by the Sun's rotation. Even though Wolszczan et
al. (2000)\nocite{whk+00} proved that the periodicity was due to
planetary companions (based on the use of the original \textsc{tempo1}
solar-wind model and multi-frequency observations) it is of interest
to understand the effect of an unmodelled (or poorly modelled) solar wind
on the measured pulsar parameters.

Recently, \citet{ojs07} observed PSRs~J1801$-$2304, J1757$-$2421,
J1757$-$2223 and J1822$-$2256 when their lines of sight were close to
the Sun. Their work has some overlap with ours as they also used
observations from the Wilcox Solar Observatory and a model for the
solar electron density. However, they concentrated on variations in
pulsar rotation measures due to the solar magnetic field.  In our work
we model variations in pulsar dispersion measures and describe their
implications for high precision pulsar timing.

In this paper we first describe a two-state solar wind model
(\S\ref{sec:swm}) and our analysis technique (\S\ref{sec:data}) before
considering the implications for high precision pulsar timing
(\S\ref{sec:discussion}).  

\section{The two-state solar wind model}\label{sec:swm}
The solar wind is a complex system and important features are still
poorly understood. A summary of the relevent physics can be found in
\cite{sch06}. In brief, the solar wind can be thought of as having a
quasi-static component which is bimodal and co-rotates with the Sun,
and a transient component which has a time scale of hours to days. The
best known of the transient events are coronal mass ejections, which
typically cross any given line of sight about 5\% of the time
\citep{sch96b}. It is currently not feasible to model the complex
transient events and we will concentrate on modeling the co-rotating
wind structure, which has ``fast'' and ``slow'' components.

The slow wind has a relatively high density and apparently originates
in or around active regions of closed magnetic geometry at low or
middle latitudes. The fast wind has lower density and originates in
regions with open magnetic field geometry called coronal holes. Large
coronal holes are located over the solar poles during the years of
minimum solar activity. Smaller and shorter-lived coronal holes occur
at middle and low latitudes when solar activity is higher. 

We note that original \textsc{tempo1} model can be thought of assuming
that the entire wind is a spherically symmetric slow wind, whereas the
default \textsc{tempo2} model assumes that the wind is entirely fast.
 
The electron density in the fast wind can be estimated from
\emph{Ulysses} and \emph{SPARTAN} observations to give:
\begin{eqnarray}\label{eqn:nefast} 
n_e = 1.155\times10^{11}R_\odot^{-2} + 32.3\times10^{11}R_\odot^{-4.39} + \\ \nonumber
3254\times10^{11}R_\odot^{-16.25}\;\;{\rm m}^{-3}
\end{eqnarray}
at a distance of $R_\odot$ solar radii \citep{gf95a,gf98}.

We can approximate the electron density in the ``slow wind'' using a
combination of the Muhleman \& Anderson (1981)\nocite{ma81} model fit
to their own observations far from the Sun and the ``Baumbach-Allen''
model near to the Sun (Allen 1947)\nocite{all47},
\begin{eqnarray}\label{eqn:neslow} 
  n_e = 2.99\times10^{14}R_\odot^{-16} + 1.5\times 10^{14}R_\odot^{-6}
  + \\ \nonumber
  4.1\times 10^{11}\left(R_\odot^{-2} + 5.74R_\odot^{-2.7}\right)\;\;{\rm m}^{-3}.
\end{eqnarray}

\begin{figure}
 \includegraphics[width=6cm,angle=-90]{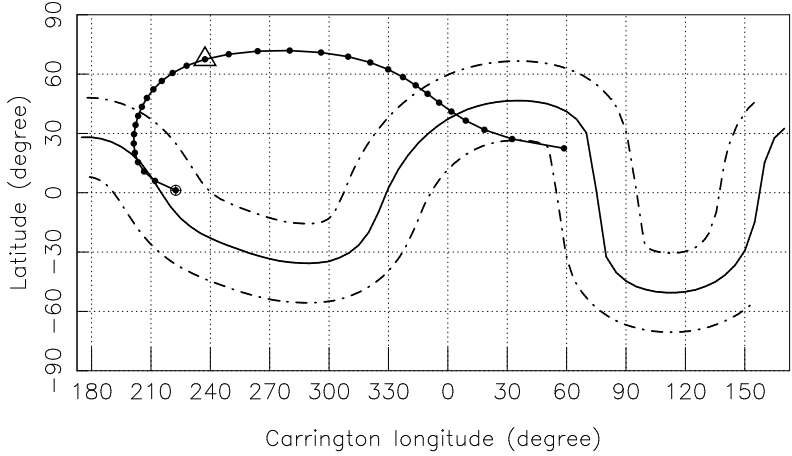}
 \caption{Projection on to the solar surface along wind streamlines of
   the line of sight to PSR~J1744$-$1134 on 2004, December 20. The
   triangle shows the point of closest approach to the Sun and the
   open circle is the projected position of the Earth. Points are at
   5$^\circ$ intervals in angle subtended at the Sun. The solid line
   indicates the position of the magnetic neutral line; the dashed
   lines on either side are plotted 20$^\circ$ away from the neutron
   line and delimit the region assumed to be dominated by the slow
   wind.}
  \label{fg:synoptic}
 \end{figure}

In order to determine DM$_\odot$ the electron density must be
integrated along the line of sight to the pulsar.  Information on
whether a given position along the line of sight will be within the
slow or the fast wind can be obtained from the Wilcox Solar
Observatory\footnote{
\url{http://soi.stanford.edu/\~{}wso/forms/prsyn.html}.  To obtain
data-sets suitable for \textsc{tempo2} the \textsc{ClassicSS} map
should be selected with a \textsc{latitude} projection. Full details
can be obtained from the \textsc{tempo2} on-line documentation
(http://www.atnf.csiro.au/research/pulsar/tempo2).  } which provides
daily maps of the solar magnetic field since May 1976. Following
\citet{mbf+00}, we assume that the slow wind occupies the zone within
20$^\circ$ of the magnetic neutral line and outside this is dominated
by the fast wind and that both winds flow radially.  To demonstrate
our technique we show, in Figure~\ref{fg:synoptic}, a synoptic chart
showing the projection of the line of sight on to the Sun for
PSR~J1744$-$1134 on the 20th December 2004.  As expected, this figure
shows that some parts of the line of sight lie within the slow wind
and some within the fast wind.

\section{Data analysis and method}\label{sec:data}
We use observations obtained for the Parkes Pulsar Timing Array (PPTA)
project \citep{man06} to test our new model.  A sample of 20
millisecond pulsars has been observed since February 2004 at intervals
of 2-3 weeks at frequencies around 700\,MHz, 1400\,MHz and
3100\,MHz. Details of the observations and the methods used to
determine the DM variations are given by \citet{yhc+07}.
 
For this paper we use data for four pulsars which have measurable DM
variations due to the solar wind.  PSRs~J1022+1001 and J1730$-$2304
have ecliptic latitudes of $-0.064^\circ$ and $0.19^\circ$
respectively and hence are eclipsed by the Sun each
year. PSRs~J1744$-$1134 and J1909$-$3744 have higher ecliptic
latitudes ($11^\circ$ and $-15^\circ$ respectively) but can be timed
with very high precision.

We have implemented algorithms into \textsc{tempo2} to integrate the
electron density along a given line of sight assuming the fast- and
slow-wind electron densities as given in Equations~\ref{eqn:nefast}
and \ref{eqn:neslow}.  For every observation, \textsc{tempo2}
calculates the projection of points along the line of sight to the
pulsar on to the surface of the Sun, assuming the Carrington rotation
rate and a mean wind velocity of 400\,km\,s$^{-1}$. These parameters
are characteristic of the slow wind and are chosen since this
component dominates both the wind dynamics and the dispersion
contribution. Using data from the Wilcox Solar Observatory,
\textsc{tempo2} determines the position of the magnetic neutral line
and, hence, the regions along the line of sight that are within the
slow and fast winds.  A numerical integration is then carried out to
obtain the total electron column density of the solar wind along the
line of sight and hence ${\rm DM}_\odot$. The derived values are not
significantly dependent upon our assumptions about the wind rotation
and velocity.

\section{Results and Discussion}\label{sec:discussion}
\begin{figure}
 \includegraphics[width=7cm,angle=-90]{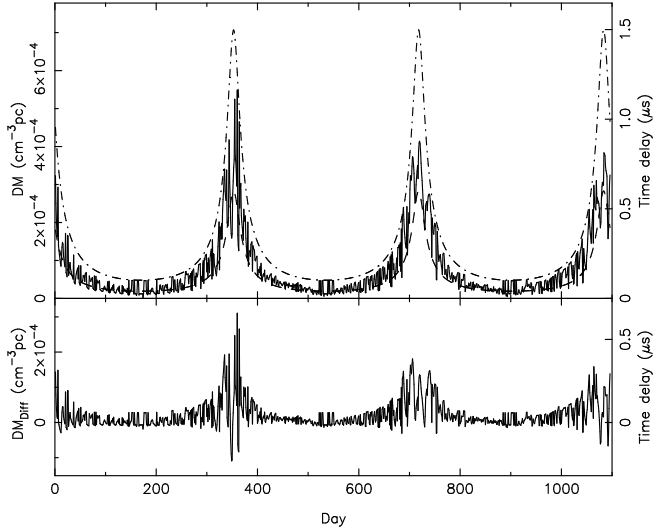}
 \caption{Solar-wind DM variations for PSR~J1744$-$1134 from 2004 to
 2006. The right-hand axis gives the corresponding time delay for an
 observing frequency of 1400\,MHz. In the upper panel the solid line
 gives the DM variations from our new model. The dashed and dot-dashed
 lines indicate predictions of the original \textsc{tempo2} model and
 the \textsc{tempo1} models respectively. In the lower panel we plot
 the difference between the new model and the original \textsc{tempo2}
 model.}
  \label{fg:DMs}
\end{figure}

Tools are available within the \textsc{tempo2} software to obtain such
synopotic charts for any pulsar on any day since the start of the
Wilcox Solar Observatory data in 1976. Figure~\ref{fg:DMs} shows the
DM variations according to our solar wind model for PSR~J1744$-$1134
between the years 2004 and 2006. The predictions according to the
earlier \textsc{tempo1} and \textsc{tempo2} models are also indicated
in the Figure. We notice that the new model generally predicts DM
values that are higher than the \textsc{tempo2} model, but lower than
the \textsc{tempo1} model as expected from our two-state model. The
new model is also not smooth.  Variations of up to DM$_\odot \sim
10^{-4}$\,cm$^{-3}$pc occur on a daily basis.
 
\begin{figure*}
  \includegraphics[width=8cm]{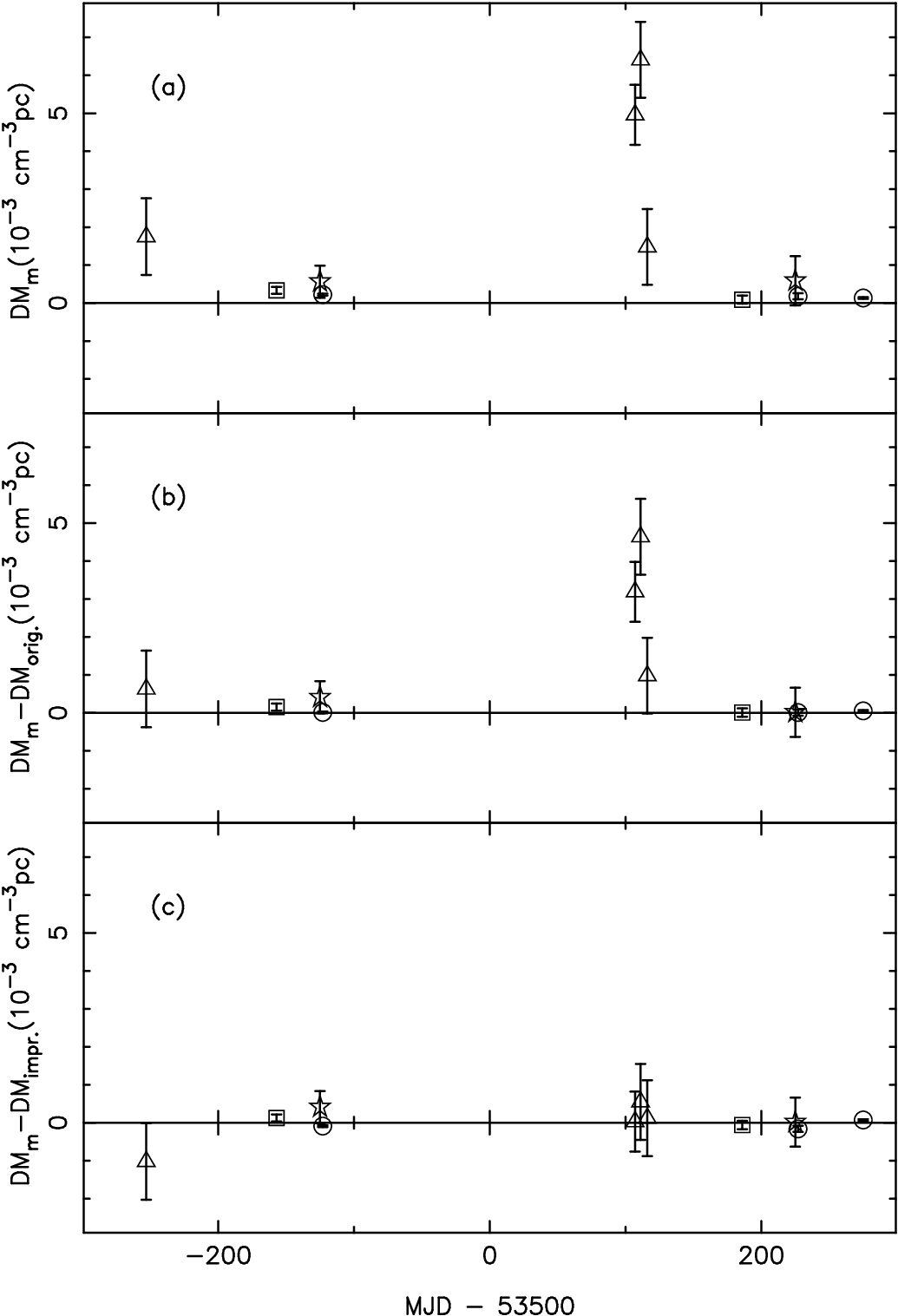}
  \includegraphics[width=8cm]{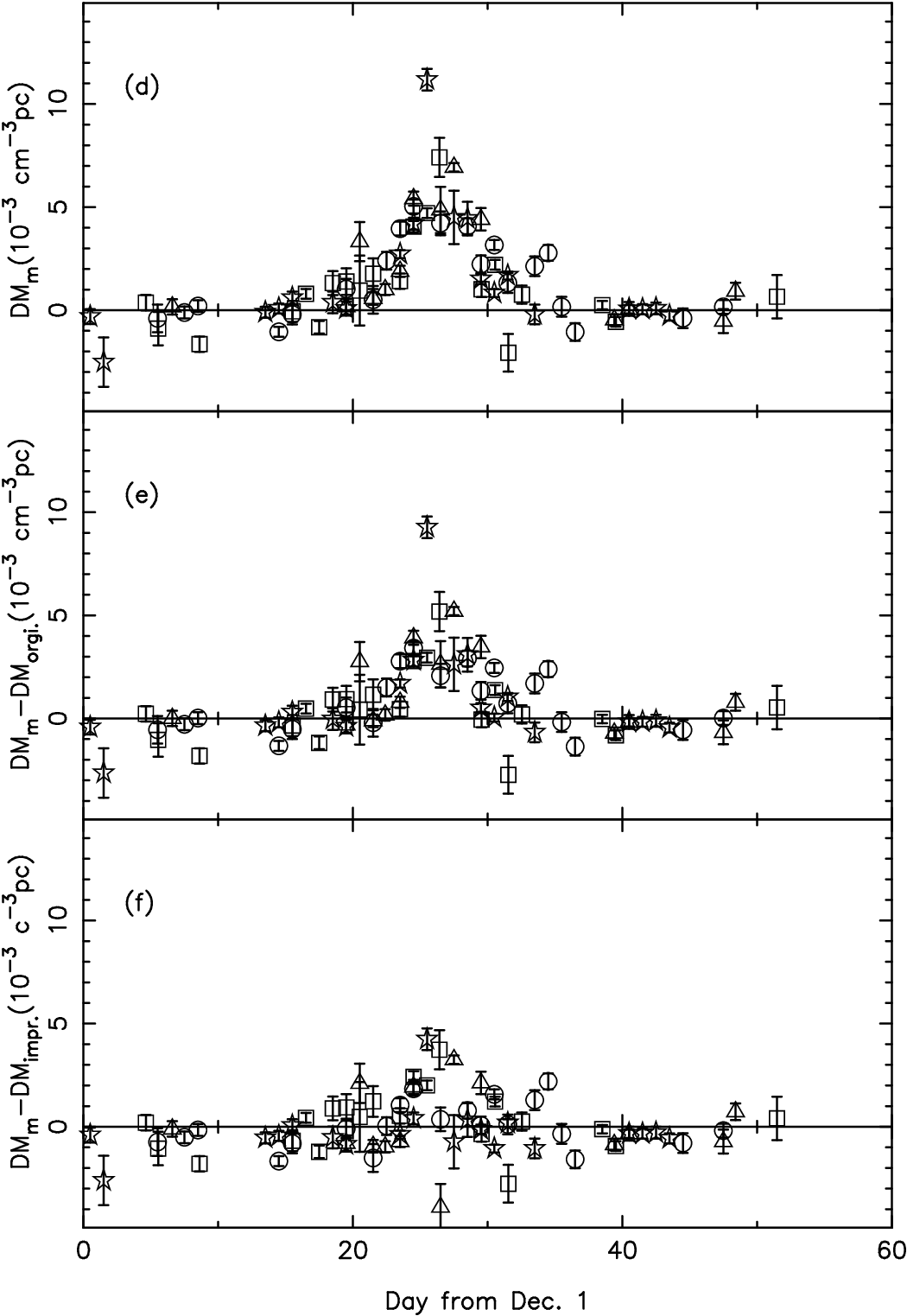}
  \caption{Left column: comparison of the measured DM values with the
  model predictions for PSRs~J1022+1002 (triangle symbols),
  J1730$-$2304 (open stars), J1744$-$1134 (open squares) and
  J1909$-$3744 (open circles).  Right column: comparison of the
  measured and predicted DMs for the Cognard et al. (1996)
  observations of PSR~B1821$-$21. Triangle, star, square and circle
  symbols represent data starting in December 1989, 1990, 1991 and
  1992, respectively.  In both columns the upper panels gives the
  measured DM values without any solar wind correction. The middle
  panels give the difference between the actual values and those
  predicted using the original \textsc{tempo2} model.  The lowest
  panels show the difference between the measurements and the
  prediction using the improved solar wind model.}\label{fg:compare}
\end{figure*}

Differences between measured DM$_\odot$ values \citep{yhc+07} and the
predictions using the original and new \textsc{tempo2} models are
shown in Figure~\ref{fg:compare} for the four pulsar datasets
discussed in this paper. This figure shows that, for lines of sight
that pass close to the Sun, the original \textsc{tempo2} does not
correctly predict DM$_\odot$.  However, the improved model predicts
DM$_\odot$ within experimental uncertainties for all observations.

Our data-sets are currently poorly sampled for lines of sight that
pass close to the Sun.  We have, therefore, compared our improved
model predictions with observations of PSR~B1821$-$24 using the Nan{\c
c}ay radio telescope \citep{cbl+96}.  DM values were measured from
their Figure~7. There appear to be significant non-solar variations in
their measured DMs and we have removed a straight line fitted to the
values more than $\sim 40 {\rm R}_\odot$ from the Sun. Comparisons
with the original \textsc{tempo2} model and our improved model are
shown in the right hand column of Figure~\ref{fg:compare}. It is clear
that the new model is a significant improvement over the previous
\textsc{tempo2} model. However, even with our improved model there are
some observations that are not consistent with our predictions.  These
inconsistencies occur at the closest approach of the line of sight to
the Sun.  At such close approaches our simple assumptions of the
two-component wind model where the slow wind lies within 20$^\circ$ of
the magnetic neutral line and radial wind flow with projection along
mean flow streamlines may not be valid.  We will be able to further
test our model with future PPTA datasets having more precise and more
closely sampled DM measurements at close angular distances to the
Sun. Such results should help us to further improve the model.

\subsection{Implications for high precision pulsar timing}
Modern pulsar timing experiments are aiming to achieve rms timing
precisions close to 100~ns over many years. At an observing frequency
of 1.4~GHz the solar wind causes time delays of this magnitude for
pulsars up to 60$^\circ$ from the Sun and significant deviations
between the original and improved \textsc{tempo2} models occur at
$\lesssim 20^\circ$ from the Sun.

In order to study the effect of an unmodelled, or poorly modelled,
solar wind on pulsar timing parameters we used \textsc{tempo2} to
create simulated data-sets spanning three years for
PSR~J1744$-$1134. For these simulations we applied the improved
solar-wind model and a specified amount of uncorrelated pulsar timing
noise. We then either switched off all solar wind models or used the
original \textsc{tempo2} model before fitting for the pulsar's
parameters. Deviations from the true values for various astrometric
parameters are listed in Table~\ref{tb:diffParams}.  Clearly, the
solar-wind model has a large effect on the values of the fitted
parameters.  For instance, for 100\,ns rms timing residuals, the
derived values for parallax and declination when using the standard
\textsc{tempo2} model deviate by $\sim 2.5\sigma$ and $\sim
11.0\sigma$ respectively from their true values.  For any given
pulsar, the error in each parameter will depend upon the rms timing
residual, the data-span and the ecliptic latitude of the pulsar.

\begin{table}\caption{Effect on timing parameters for different rms timing 
residuals when comparing the new solar wind model to 1) no model and
2) the original \textsc{tempo2} model.}
\label{tb:diffParams}
\begin{center}\begin{tabular}{clrr}\hline
Rms resid. & \multicolumn{1}{c}{Parameter} & No model & Orig.T2
model\\ ($\mu s$) & & \multicolumn{1}{c}{($\sigma$)}
&\multicolumn{1}{c}{($\sigma$)} \\ \hline 0.0 & Right ascension & 11.8
& 9.0 \\ & Declination & 52.1 & 22.2 \\ & Parallax & 26.6 & 4.1 \\ 0.1
& Right ascension & 5.9 & 2.7 \\ & Declination & 33.4 & 11.0 \\ &
Parallax & 17.6 & 2.5 \\ 1.0 & Right ascension & 1.3 & 1.7 \\ &
Declination & 3.8 & 0.8 \\ & Parallax & 2.7 & 0.7 \\ \\ \hline
\end{tabular}\end{center}
\end{table}

 \begin{figure}
  \includegraphics[width=6cm,angle=-90]{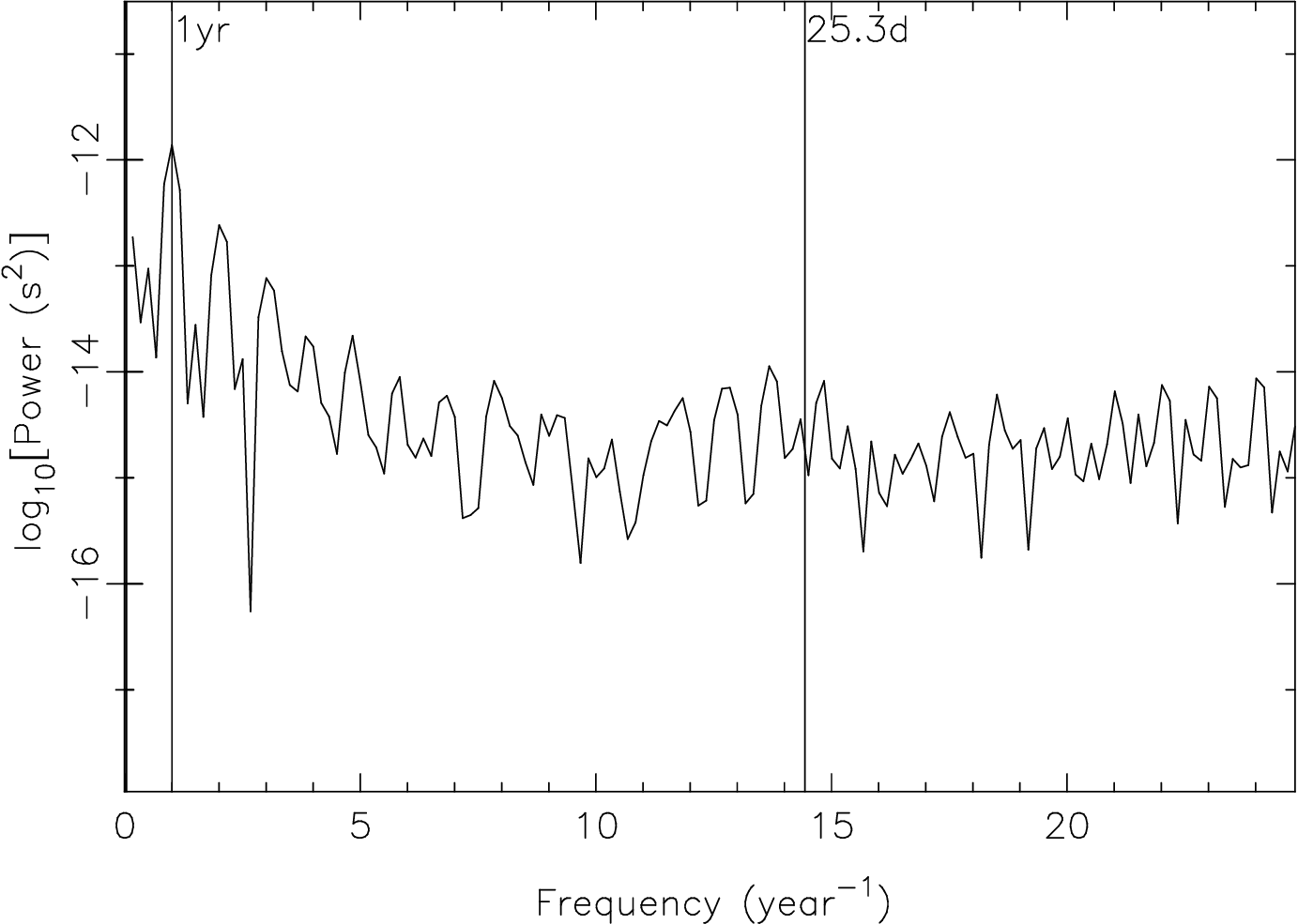}\\
  \includegraphics[width=6cm,angle=-90]{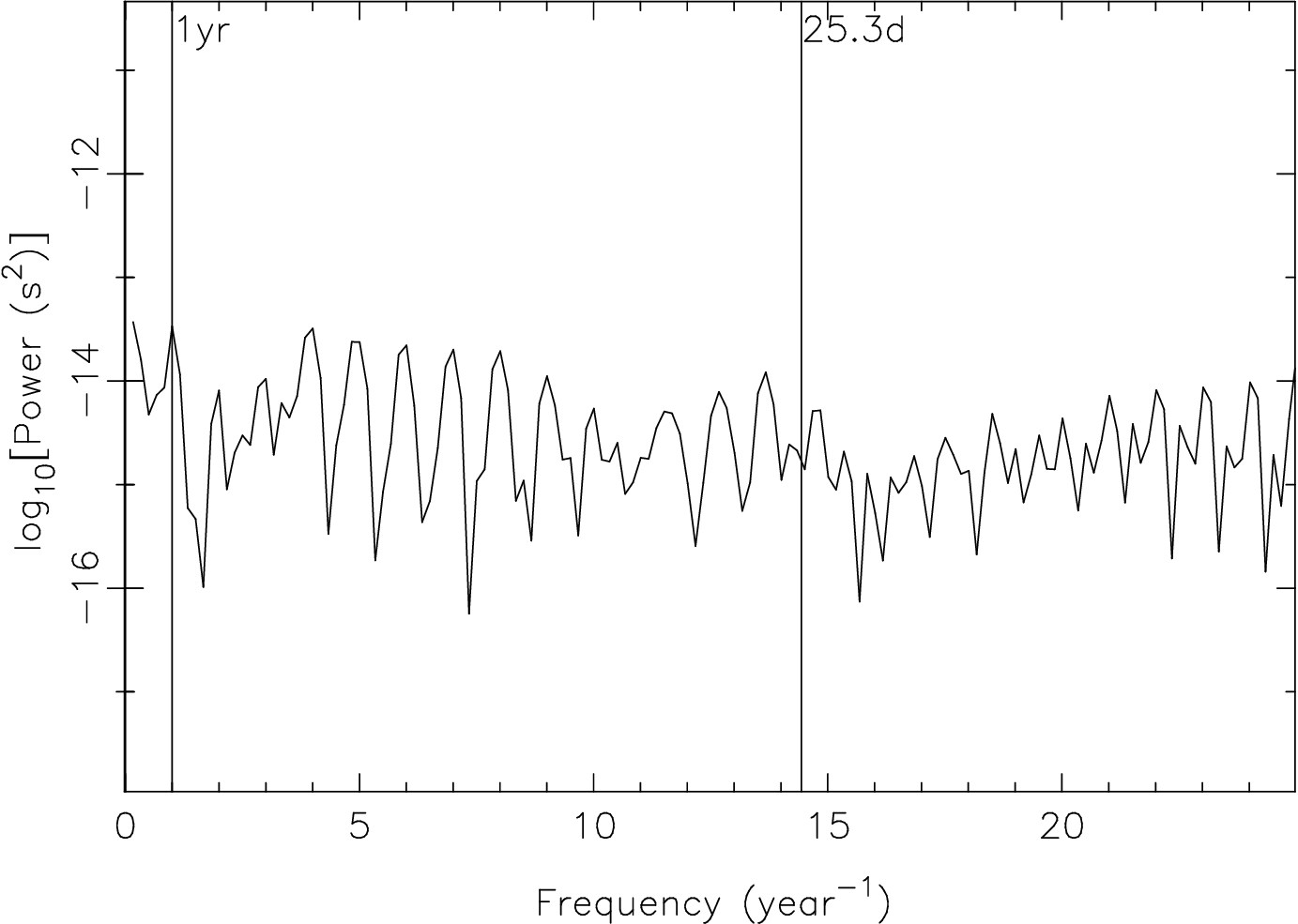}\\
  \includegraphics[width=6cm,angle=-90]{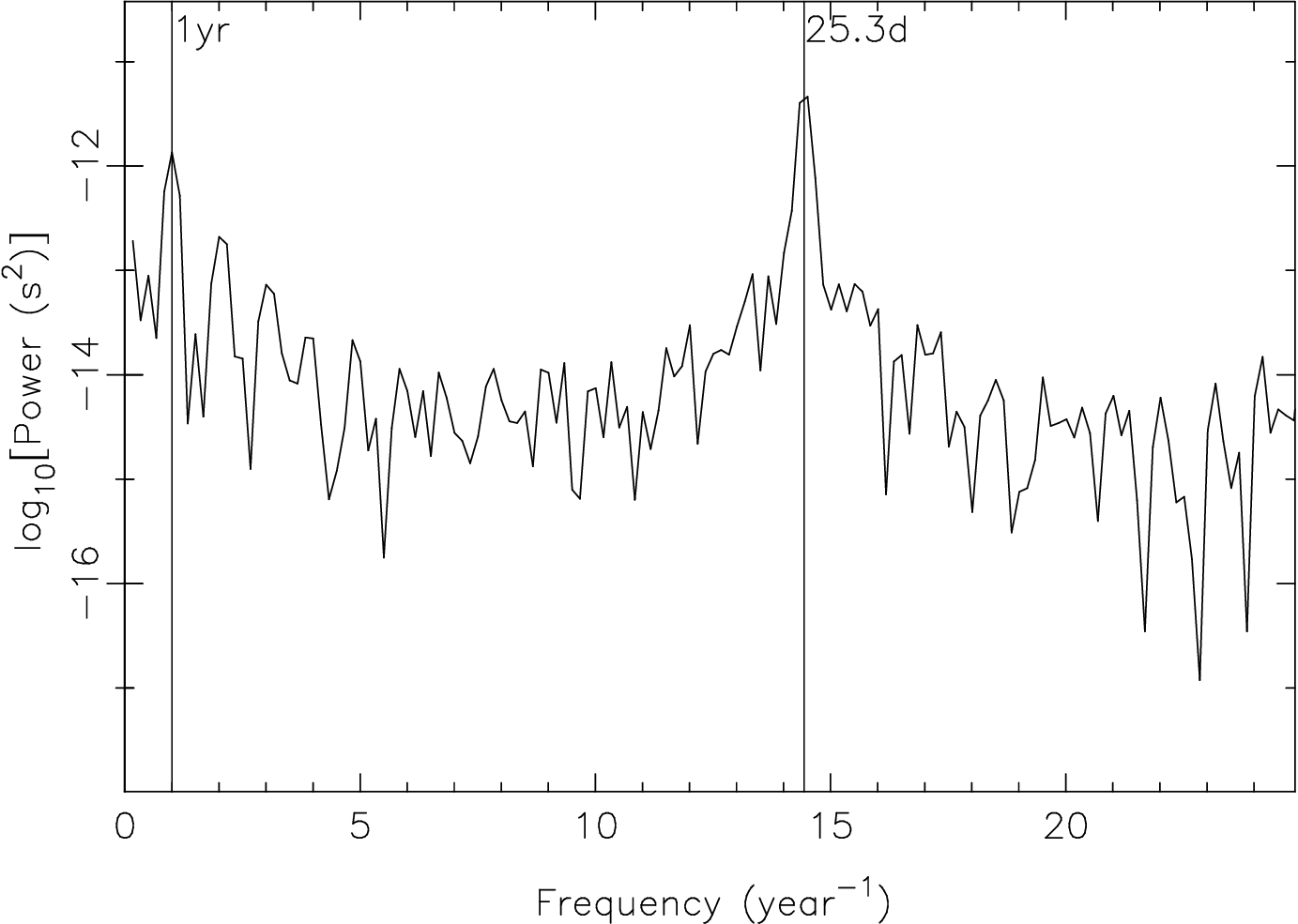}
  \caption{The upper panel shows the power spectrum of timing
  residuals contributed by the improved model of the solar wind for a
  simulated three-year data-span for the PSR~B1257+12 system.  In the
  middle panel we have reduced the power in the annual term and its
  harmonics by fitting the spherical wind model to the residuals and
  subtracting it. The bottom panel shows the power spectrum of
  residuals due to both the solar-wind model and the inner planet of
  the PSR~B1257+12 system.}\label{fg:pspectrum}
 \end{figure}

In order to test whether the solar wind can mimic planetary companions
we have simulated a data-set for PSR~B1257$+$12 with the same span and
observing frequency as the Wolszczan (1994) observations. As we have
no access to the original Wolszczan (1994) data-set, our simulated
observations are uniformly sampled. A power spectrum of the solar wind
contribution was computed with a rectangular window using the
Lomb-Scargle algorithm. Since the data are uniformly sampled this is
the same as the normal Fourier power spectrum. This spectrum, shown in
the top panel of Figure~\ref{fg:pspectrum}, is dominated by harmonics
of the annual modulation. To reduce the leakage of annual harmonics
into the higher frequencies we removed the annual feature by
subtracting a fit of a spherically symmetric model
(Equation~\ref{eqn:model}) to the residuals. The resulting spectrum is
shown in the middle panel. There is no significant feature in either
spectrum corresponding to the narrow 25.3-d peak seen in solar wind
observations by Scherer et al. (1997). In fact we do not expect to see
a sharp feature in the spectrum because the line of sight to the
pulsar changes significantly during a solar rotation and the solar
wind density itself evolves on that time scale. We have used
\textsc{tempo2} to introduce the expected signal from the planet and
applied the same spectral analysis to the combined solar wind plus
planet simulation. The resulting spectrum is plotted in the bottom
panel. The contribution of the planet exceeds the solar wind noise by
a factor of more than 100. Clearly Scherer et al. (1997) seriously
overestimated the importance of solar noise in the detection of a
planet around PSR~B1257$+$12.

\section{Conclusions}
We have developed a new solar-wind model for pulsar timing experiments
and shown that it gives a more accurate correction for delays due to
the solar wind than earlier models. Use of the older solar-wind models
(or no correction) leads to systematic errors in measured pulsar
parameters. We have also shown that the solar wind cannot mimic the
signal from inner-most planetary companion of PSR B1257+12 as
suggested by Scherer et al. (1997). With the improved pulsar timing
data expected in the future from projects such as the Parkes Pulsar
Timing Array, use of the new model will make an important contribution
to achieving the goals of these projects. The improved model has been
implemented in the \textsc{tempo2} software package and we recommend
that it be used for all high-precision timing applications.

\section*{Acknowledgments} 

X.P.Y. and J.L.H. are supported by the National Natural Science
Foundation of China (No.10473015 \& 10521001).  W.A.C. was partially
supported by the National Science Foundation of the U.S.A. under grant
AST0507713. The Wilcox Solar Observatory data used in this study were
obtained via the web site {\url{http://soi.stanford.edu/\~{}wso}} at
\verb|2007:06:07_21:32:20 PDT| courtesy of J.T. Hoeksema.  The Wilcox
Solar Observatory is supported by NASA.  The data presented in this
paper were obtained as part of the Parkes Pulsar Timing Array project
that is a collaboration between the ATNF, Swinburne University and the
University of Texas, Brownsville, and we thank our collaborators on
this project.  The Parkes radio telescope is part of the Australia
Telescope which is funded by the Commonwealth of Australia for
operation as a National Facility managed by CSIRO.


\end{document}